\documentstyle[prb,aps,twocolumn]{revtex}
\begin{document}
\draft
\title{Theory of Orbital Excitation and \\  
Resonant Inelastic X-ray Scattering in Manganites}
\author{Sumio Ishihara and Sadamichi Maekawa}
\address{Institute for Materials Research, Tohoku University, 
Sendai 980-8577 Japan}
\date{\today}
\maketitle
\begin{abstract}
We study theoretically the collective orbital excitation named orbital wave 
in the orbital ordered manganites. 
The dispersion relation of the orbital wave is affected by the static spin structure 
through the coupling between spin and orbital degrees of freedom. 
As a probe to detect the dispersion relation, 
we propose two possible methods by utilizing resonant inelastic x-ray scattering. 
The transition probability of the orbital wave scattering is formulated, and the 
momentum and polarization dependences of the structure factor are 
calculated in several types of the orbital and spin structures. 
The elastic x-ray scattering in the L-edge case to observe the orbital ordering 
is also discussed. 
\end{abstract}
\pacs{71.10.-w, 71.90.+q, 75.90.+w, 78.70.Ck}
\narrowtext
\noindent
\section{Introduction}
It is widely accepted that one of the important ingredients for  
the colossal magnetoresistance (CMR) \cite{kaha,heru,toku,jin}
and various complex phenomena 
observed in perovskite manganites 
is the orbital degree of freedom in $\rm Mn^{3+}$ ion. 
Although the long range orbital ordering and its 
implications of the magnetic interaction were 
discussed long time ago, \cite{good,kana,kugel} 
it has been recognized as a hidden degree of freedom, 
because the observation technique has been 
limited. \cite{akimitsu}
Recently, the anomalous x-ray scattering (the resonant elastic x-ray scattering) 
shed light on the subject, that is, 
by using the method the orbital ordering in 
$\rm La_{0.5}Sr_{1.5}MnO_4$ was directly observed. \cite{mura1}
The alternate orbital alignment was confirmed by observation of the forbidden reflection 
originated from orbital dependence of the atomic scattering factor. 
In a moment, this experimental technique was recognized as a 
powerful probe to detect the orbital structure and  
applied to several undoped and doped manganites. \cite{mura2,mura3,arima1}
Through the intensive experimental and theoretical studies, \cite{ishi4,ishi5,fab}
it certainly uncovers roles of the orbital ordering on the electronic 
and lattice structures in manganites. 
\par
On the other hand, the dynamics of the orbital degree of freedom is still far 
from our understanding . 
In orbital ordered insulators, 
the collective excitation for the orbital degrees was  
first investigated by Cyrot {\it et al.} in the model where the 
orbital space was assumed to be isotropic. \cite{cyro}
It was termed orbital wave in analogy with spin wave in a  
magnetic ordered state. 
For actual compounds, 
the present authors calculated the dispersion relation of the 
orbital wave in the $(3d_{3x^2-r^2}/3d_{3y^2-r^2})$-type orbital 
ordered state with A(layer)-type antiferromganetic spin structure 
observed in $\rm LaMnO_3$. \cite{ishi1}
When such a collective excitation in the orbital degree exists, 
it probably affects the thermodynamics and the dispersion relations of 
spin wave \cite{brink1} and phonon. 
Furthermore, 
the orbital excitation in the metallic phase 
seems to play an essential role in the transport and 
optical properties \cite{ishi2} and CMR. 
As a probe to 
observe the orbital excitation, 
Inoue {\it et al}. theoretically  examined 
the Raman scattering \cite{inou}
where the excitation process is analogous of that in the magnon Raman scattering. 
In this method,  
the information of the $\Gamma$-point and the density of state 
of the orbital wave are obtained. 
\par
In this paper, we study theoretically the orbital wave in 
orbital ordered manganites and propose a method  
to observe it by the resonant inelastic x-ray scattering (RIXS). 
RIXS is rapidly developed 
through the recent progress of the synchrotron radiation sorce \cite{rixs}
and is applied to several highly correlated electron 
systems. \cite{kuiper,hill}  
One of the advantages of this method on the subject 
is that the wave length of the x-ray is comparable to the 
lattice constant, unlike the conventional light. 
Therefore, the dispersion relation of the orbital wave 
in the wide range of the Brilloun zone 
is detectable by the method. 
The following two different processes in RIXS are proposed: 
the incident photon energy is tuned 
at 1) $\rm Mn^{3+}$ L-edge and 2) $\rm Mn^{3+}$ K-edge. 
We formulate the transition probability in 
both types of the scattering and the structure factor 
is calculated as a function of momentum and 
polarization in several orbital orderings. 
\par
In Sec. II, 
the dispersion relation of the orbital wave is calculated. 
We emphasize the strong correlation between the static spin structure 
and the orbital wave. 
In Sec. III, 
a mechanism to observe the orbital wave by RIXS is introduced. 
The general formulas of the scattering intensity and the numerical results of the 
structure factor are presented. 
Sec. V is devoted to the discussion and summary. 
\section{orbital wave}
In order to calculate the 
collective excitations in an orbital ordered insulator, 
we start with the Hamiltonian 
describing the low energy 
electronic structure in perovskite manganites. 
The cubic lattice consisting of Mn ions 
is considered and two $e_g$ orbitals and 
a $t_{2g}$ localized spin are introduced in 
each site. 
The Coulomb interactions 
between $e_g$ electrons $(U,U',J)$ and 
the Hund coupling $(J_H)$ between $e_g$ and $t_g$ 
spins are considered. 
Because the Coulomb interaction
gives the largest energy 
among the relevant parameters, 
the Hamiltonian is derived by excluding 
the doubly occupied $e_g$ states as follows, 
\begin{equation}
\widetilde H_{3d}=H_{e \mbox{-} e}+H_{e \mbox{-} t}+H_{t \mbox{-} t} \ . 
\label{eq:hamiltonian}
\end{equation}
The detailed derivation is presented in Ref.\onlinecite{ishi1}. 
The first term describes the interaction 
between nearest neighboring spins and orbitals in $e_g$ orbitals, 
\begin{eqnarray}
H_{e \mbox{-} e}=&-&2J_1\sum_{\langle ij \rangle } 
\Bigl ( {3 \over 4} + \vec S_i \cdot \vec S_j   \Bigr )
\Bigl ( {1 \over 4} - \Psi_i^\dagger \hat {\tau}_{ij} \Psi_j \Bigr ) \nonumber \\
     &-&2J_2\sum_{\langle ij \rangle } 
\Bigl ( {1 \over 4} - \vec S_i \cdot \vec S_j   \Bigr )
\Bigl ( {3 \over 4} + \Psi_i^\dagger \hat {\tau}_{ij} \Psi_j + A_{ij} \Bigr ) \ ,
\label{eq:effect}
\end{eqnarray}
where a relation $U=U'+J$ is assumed 
with $U$, $U'$, and $J$ being the intra- and inter-orbital 
Coulomb interactions and the exchange interaction in $e_g$ 
orbitals, respectively. 
$J_{1}=t_0^2/(U'-J)$ and $J_{2}=t_0^2/(U'+J+2J_H)$ 
with $J_1 > J_2$. 
$ \Psi_i=\bigr [ T_{iz}, T_{ix } \bigl ]^{t } $ 
is a doublet of the orbital pseudo-spin operator:
$\vec T_{i}=(1/2) 
\sum_{\sigma \gamma \gamma'}
\widetilde d_{i \gamma \sigma}^\dagger (\vec \sigma)_{\gamma \gamma'}
\widetilde d_{i \gamma' \sigma} \ $ 
and $\vec S_{i \gamma_\theta}$ is the spin operator with $S=1/2$ 
for an $e_g$ electron of orbital $\gamma_\theta$.  
$\hat \tau_{ij}$ and $A_{ij}$  
are defined by 
\begin{equation}
\hat \tau_{i i+l}= { 1 \over 2}
\pmatrix { 1+\cos n_l {2 \pi \over 3}
       &  \sin n_l {2 \pi \over 3} \cr 
\sin n_l {2 \pi \over 3}  
     &   1-\cos n_l {2 \pi \over 3}
 \cr } \ , 
\end{equation}
and 
\begin{equation}
A_{ij}=\Psi^\dagger_i M_{i j} n_j+H.c. \ , 
\label{eq:AAAA}
\end{equation}
respectively, 
where $M_{i i+l}=( 
\cos n_l {2 \pi \over 3}, -\sin n_l {2 \pi \over 3})^t $ 
with $(n_x,n_y,n_z)=(1,2,3)$. 
The forms of 
$\hat \tau_{ij}$ and $M_{ij}$ 
in the above are 
obtained by the Slater-Koster formulas 
for the transfer intensity 
$t_{ij}^{\gamma \gamma'}$ 
between site $i$ with orbital $\gamma$ and 
site $j$ with $\gamma'$.  
It is worth noting that 
the orbital part of $H_{e \mbox{-} e}$ is represented by $T_z$ and $T_x$, 
since $t_{i j}^{\gamma \gamma'}$ is rewritten by the rotation matrix of the 
$(T_z,T_x)$-plane and the projection operator for $3d_{3z^2-r^2}$ orbital. 
The first term in Eq.(\ref{eq:effect}) 
favors the  ferromagnetic spin structure with the 
antiferro-type orbital ordering 
where two kinds of orbital sublattice exist.
On the other hand, 
the second one favors 
the antiferromagnetic 
structure with the ferro-type orbital ordering 
where occupied orbital is uniform. 
The sum of the 
second and third terms in Eq.(\ref{eq:hamiltonian}) is 
given by 
\begin{eqnarray}
H_{e \mbox{-} t}+H_{t \mbox{-} t}
=&-&J_H \sum_{i \gamma_\theta} \vec S_{t_{2g} i } 
\cdot \vec S_{i \gamma_\theta} 
\nonumber \\
&+& J_{AF} \sum_{\langle i j \rangle} \vec S_{t_{2g} i} 
\cdot  \vec S_{t_{2g} j} \ ,  
\label{eq:hhund}
\end{eqnarray}
where $J_{AF}$ is the antiferromagnetic superexchange interaction 
between nearest neighboring $t_{2g}$ spins $\vec S_{t_{2g} i }$
with $S=3/2$. 
\par
We first study the spin and orbital 
ordered structures at zero temperature in the 
mean field approximation. 
It is assumed that 
in each Mn site, one of the $e_g$ orbitals 
is occupied by an electron. 
For both spin and orbital structures, 
we introduce four-types of the ordering: 
ferro-type (F-type) where spins or orbitals in all 
sites are parallel, and three kinds of antiferro-type, 
that is, layer-type (A-type), rod-type (C-type), and 
NaCl-type (G-type) where two kinds of spin or orbital  
sublattice exist. 
The rotating frame is introduced in the orbital space and 
the orbital state is described by an angle $\theta$ in the $(T_z,T_x)$-plane. 
In the scheme, the occupied orbital is represented by 
$|3d_{\gamma_\theta} \rangle=
\cos(\theta/2) |3d_{3z^2-r^2} \rangle -
\sin(\theta/2) |3d_{x^2-y^2} \rangle$.
As the order parameters, $\langle S_z \rangle $, 
$\langle S_{t_{2g} z} \rangle $ and 
$\langle T(\theta)_z \rangle
=\cos \theta \langle T_z \rangle
-\sin \theta \langle T_x \rangle$ 
with $\langle S_{t_{2g} z} \rangle =3\langle S_z \rangle $ are introduced. 
\par
In Fig. 1, the mean field phase diagram is presented. 
A sequential change in the spin structure, that is, $F \rightarrow A \rightarrow 
C \rightarrow G$ with increasing $J_2$ and $J_{AF}$ 
is caused by an enhancement of the antiferromagnetic 
interaction between spins. 
As for the orbital structure 
in the spin-F and spin-G phases, 
the orbital-G with $(\theta_A/\theta_B)=(\theta_A/\theta_A+\pi)$ 
for any $\theta_A$ 
and the orbital-C with $(\theta_A/\theta_B)=({\pi \over 2} / { 3\pi \over 2})$ 
are the mean field solutions \cite{kugel,maezo,shii} with 
$\theta_A$ and $\theta_B$ being the angles in $A$ and $B$ orbital 
sublattices, respectively.  
In the orbital-G case, especially, 
the orbital space becomes isotropic in the $(T_z,T_x)$-plane, 
although the orbital part in the Hamiltonian 
Eq.(\ref{eq:effect}) is not written by a simple vector product. 
This result is originated from 1) the relation,   
\begin{equation}
\sum_{l=x,y,z} t_{i i+l}^{\gamma \gamma'} =
 {3 \over 2} t_0 \delta_{\gamma \gamma'} \ , 
\label{eq:tcond}
\end{equation}
due to the cubic symmetry of the lattice 
and 2) the cubic symmetry of the spin structure, that is, spin-F and -G. 
Because of these two conditions, 
the orbital part in $H_{e \mbox{-} e}$ becomes diagonal  
and $A_{ij}$ term vanishes.  
On the other hand, in spin-A and -C phases, 
the orbital state is uniquely determined. 
In spin-A, the orbital structure in the mean filed theory is 
orbital-C and -G types with $(\theta_A/\theta_B)= (\theta_A/-\theta_A)$ where 
$\theta_A$ is determined by 
the equation $\theta_A=\cos^{-1} (2J_2/(5J_1-J_2))$. 
It includes the 
$( {1 \over \sqrt{2}} 
   (3d_{3z^2-r^2}-3d_{x^2-y^2})/
 {1 \over \sqrt{2}} 
   (3d_{3z^2-r^2}+3d_{x^2-y^2}) )$ orbital order  
at $J_2=0$ and 
$(3d_{z^2-x^2}/3d_{y^2-z^2})$ order at $J_2=J_1$. \cite{kugel,maezo,shii}
The $(\theta_A/-\theta_A)$-type orbial order in spin-A 
is consistent to the experimental results 
of the polarization dependence of the resonant elastic x-ray 
scattering, \cite{mura2,ishi5} although the value of $\theta_A$ seems to be 
modified by the coupling with lattice. 
In spin-C, the mean field solution is 
orbital-G type with $(0/\pi)$ 
corresponding to the $(3d_{3z^2-r^2}/3d_{x^2-y^2})$. 
In both cases, the spin structure breaks the cubic symmetry in the system 
and causes the anisotropic interaction between nearest neighboring orbital 
pseudo-spins.  
This noticeable feature in the interplay between spin and orbital 
degrees clearly reflects on the 
dispersion relation of the orbital excitation, as will be discussed below.  
\par
We next study the orbital excitation in the orbital ordered states obtained in the 
mean field theory. 
The collective excitations in the spin and orbital degrees of freedom are 
obtained by utilizing the conventional 
Holstein-Primakoff approximation. \cite{kaliullin,feiner,brink2}
The dispersion relation in the spin-F case is shown 
in Fig. 2 where the types of the orbital order are chosen as 
$\theta_A=\pi n/6=\theta_B-\pi$. 
The face-centered cubic (fcc) lattice is adopted as a unit cell which 
includes two Mn ions. 
The analytic form of the dispersion relation 
and the eigen-operator  
are given by  
\begin{equation}
\omega^{(\pm)}_k=3\sqrt{ \alpha (\alpha \pm \beta) } \ , 
\label{eq:disf}
\end{equation}
with 
\begin{equation}
\alpha=-2J_1 {1 \over 3}  \sum_{l=x,y,z} (\tau_{i i+l})_{11} \ , 
\end{equation}
\begin{equation}
\beta=2J_1 { 1 \over 3} \sum_{l=x,y,z} \cos(ak_l) (\tau_{i i+l})_{22} \ ,  
\label{eq:alpha2}
\end{equation} 
and 
\begin{eqnarray}
\phi^{(\pm)}&=&\cosh \theta_{ k }^{(\pm)} {1 \over \sqrt{2}} (a_k \pm b_k) 
\nonumber \\
              &+&\sinh \theta_{ k }^{(\pm)} {1 \over \sqrt{2}} 
            (a_{-k}^\dagger \pm b_{-k}^\dagger)  \ , 
\label{eq:eigen}
\end{eqnarray}
with $2\theta_{ k}^{(\pm)}=\tanh^{-1} (\pm \beta /( 2\alpha \pm \beta)) $, 
respectively. 
$a_k$ and $b_k$ are the Fourier transform of the Holstein-Primakoff bosons 
for the two orbital sublattices 
defined in the rotating frame. 
The $(+)$-mode is the gapless mode 
and its eigen-operator includes the components $T_{A x}-T_{B x}$ 
and $T_{A y}+T_{B y}$. 
At the $\Gamma$-point, 
the $x$-component of the staggered orbital order parameter  
$\langle (\sum_{ i \in A } T_{i x}-\sum_{ i \in B } T_{i x})^2  \rangle $
diverges 
since the orbital space is isotropic. 
With increasing $\theta_A$, the stiffness of the orbital wave 
along the $\Gamma-X$ direction becomes weak and at $\theta_A=4 \pi/6$ 
$(3d_{3x^2-r^2}/3d_{y^2-z^2})$ 
it becomes flat. 
It is originated from $(\tau_{i i+x})_{22}=0$ in Eq.(\ref{eq:alpha2}) 
since the electron transfer between $3d_{y^2-z^2}$ orbitals along the $x$ direction 
vanishes. 
The similar gapless mode and the changes of the 
stiffness with changing types of the orbitals are also obtained in the spin-G case. 
The isotropic nature in the orbital space in spin-F case 
is seen by the Fourier transforming of the Hamltonian (Eq.(\ref{eq:effect})) as 
\begin{equation}
H_{e \mbox{-} e}=-2J_1 \sum_{\vec k} 
\Bigl( {3 \over 2}-\sum_{l=z,x} 
\widetilde T_{l}(\vec k) \widetilde \tau_l(\vec k) \widetilde T_{l}(\vec k) \Bigr) \ , 
\label{eq:effek}
\end{equation}
in the spin-F case. \cite{ishi2}
$\widetilde T_l(\vec k)$ is the orbital pseudo-spin operator 
which diagonalizes the matrix $\hat \tau_{ij}$ and 
\begin{eqnarray}
\widetilde \tau_{l}(\vec k)&=&2 \bigl(
(c_x+c_y+c_z) \cr
&\pm& (c_x^2+c_y^2+c_z^2-c_x c_y-c_y c_z-c_z c_x)^{1/2}
\bigr) \ , 
\end{eqnarray}
where $+$ and $-$ are for $l=z$ and $x$, respectively, 
and $c_l=\cos(ak_l)$. 
At the $\Gamma$- and $R$-points, $\widetilde \tau_x=\widetilde \tau_z$, 
so that the orbital system becomes isotropic. 
\par
The orbital excitation in the spin-A case 
is qualitatively different from that 
in the spin-F case as shown in Fig. 3. 
The  unit cell with four Mn ions 
is adopted, so that there exist four modes.  
Two of them have the dispersion relations, 
\begin{equation}
\omega^{(+ \pm)}_k= {1 \over 2}
\sqrt{ (2 \alpha_{xy}+ \alpha_{zz} )  
\bigl [ 2 (\alpha_{xy}+\beta_{xy})+(\alpha_{zz} \pm \beta_{zz}) 
\bigr ] } \ ,  
\label{eq:disa1}
\end{equation}
and the other two $\omega^{( - \pm )}$ are given by 
replacing $(\alpha_{xy}+\beta_{xy})$ in Eq.(\ref{eq:disa1}) by 
$(\alpha_{xy}-\beta_{xy})$.  
Here, 
\begin{eqnarray}
\alpha_{lm}=&-&2J_1 
\Bigl ( {3 \over 4}+K_{lm} \Bigr ) (\tau_{lm})_{11}
\nonumber \\
&+&2J_2
\Bigl ( {{1 \over 4}}-K_{lm} \Bigr ) ((\tau_{lm})_{11}+2(M_{lm})_{1}) \ , 
\end{eqnarray}
and 
\begin{eqnarray}
\beta_{lm}=\Bigl [ 
2J_1 \Bigl ( {3 \over 4}+K_{lm}  \Bigr )-2J_2 \Bigl ( {1 \over 4}-K_{lm}  \Bigr ) 
\Bigr ] (\gamma_{lm})_{22} \ , 
 \end{eqnarray}
for $(lm)=(xy)$ or $(zz)$. 
In these formulas, 
$A_{lm}=(1/2) \sum_{n=l,m} A_{i i+n}$ for 
$A=\tau$ and $M$, 
$\gamma_{lm}=(1/2)  \sum_{n=l,m} \tau_{i i+n}\cos(ak_n) $ 
and 
$K_{lm}$ is the spin correlation function given by 
$K_{lm}=(1/2) \sum_{n=l,m} \langle (S_{i})_z (S_{i+n})_z \rangle$. 
Among the four modes, $(++)$-mode is the lowest one and its 
eigen-operator is expressed as  
\begin{eqnarray}
\phi^{(++)}_k&=& \cosh\theta_k^{(++)} 
{1 \over 2} (a_{1 k}+a_{2 k}+b_{1 k}+b_{2 k}) 
\nonumber \\
&+& \sinh\theta_k^{(++)} {1 \over 2}
(a_{1 -k}^\dagger+a_{2 -k}^\dagger+b_{1 -k}^\dagger+b_{2 -k}^\dagger) \ , 
\label{eq:eig2}
\end{eqnarray}
where 
the subscripts 1 and 2 imply the spin sublattices. 
The operator includes the components 
$T_{A_1 x}+T_{A_2 x}-T_{B_1 x}-T_{B_2 x}$ 
and $T_{A_1 y}+T_{A_2 y}+T_{B_1 y}+T_{B_2 y}$ 
and it corresponds to $(+)$-mode in the spin-F case. 
At the $\Gamma$-point, the energy is obtained as 
\begin{equation}
\omega^{(++)}_{k=\Gamma} = { 1 \over 2}
\sqrt{ (2 \alpha_{xy}+ \alpha_{zz} ) A^{(++)} } \ ,  
\label{eq:wpp}
\end{equation}
with 
\begin{eqnarray}
A^{(++)} &=&\sum_{(lm)=(xy),(zz)} z_{lm}
\nonumber \\
\times
\Biggl \{ \Bigl [
&-&2J_1 \Bigl ( {3 \over 4}+K_{lm} \Bigr ) 
+2J_2 \Bigl ( {1 \over 4}-K_{lm} \Bigr )  \Bigr ]
\nonumber \\
&\times&
((\tau_{lm})_{11}-(\tau_{lm})_{22}) \nonumber \\
&+&2J_2 \Bigl ( {1 \over 4} -K_{lm} \Bigr ) 2 (M_{lm})_1 
\Biggr \} \ , 
\label{eq:gap}
\end{eqnarray}
where $z_{xy}=2$ and $z_{zz}=1$. 
This mode is gapful in contrast with the spin-F case. 
The right hand side in Eq. (\ref{eq:gap}) is 
represented by a product of the spin correlation 
function and the anisotropic transfer intensity. 
Therefore, the origin of the gap is attributed to the fact: 
1) 
The orbital space is anisotropic, i.e., 
$(\tau)_{11}\ne(\tau)_{22}$ 
and $A_{ij} \ne 0$, which 
originate from the 
anisotropic transfer intensity 
$t_{ij}^{\gamma \gamma'} \ne \delta_{\gamma \gamma'}$. 
2) 
The spin correlation functions 
in the $xy$ plane and in the $z$ direction are 
different ($K_{xy}\ne K_{zz}$). 
The gapful excitation is also obtained in the spin-C case. 
It is concluded that the anisotropic spin structure 
breaks the cubic symmetry in the system and causes 
the anisotropy in the orbital space, 
which causes the gap in the orbital wave. 
\section{resonant inelastic x-ray scattering }
In this section, we propose a method to detect the orbital wave by RIXS.   
The following two excitation processes are considered: 
1)  the incident energy is tuned at $\rm Mn^{3+}$ L-edge and 2) at $\rm Mn^{3+} $K-edge. 
Hereafter, the former and latter processes are termed L-edge and K-edge cases, respectively. 
The schematic processes are presented 
in Fig. 4. 
In the L-edge case, the incident x-ray excites an electron from Mn $2p$ orbital to the unoccupied Mn $3d$ orbital, and then 
one of the $3d$ electrons fills the core hole with emitting the x-ray. 
When the occupied orbitals are different between the initial and final states, 
the orbital excitation is brought about and the excitation 
propagates through the interaction between the nearest neighboring pseudo-spins. 
This process is denoted by, 
\begin{equation}
| 3d_\gamma^1 \rangle +h\nu  \rightarrow 
| 3d_\gamma^1 3d_{\gamma'}^1 \underline{2p} \rangle \rightarrow 
| 3d_{\gamma''}^1 \rangle +h\nu' \ , 
\label{eq:lprocess}
\end {equation}
where $\underline{2p}$ implies that one hole occupies the Mn $2p$ 
orbital.
In the K-edge case, the excitation occurs 
through the screening process. 
At first, the incident x-ray excites an electron from Mn $1s$ orbital 
to the unoccupied Mn $4p$ one at $i$ site. 
At the time, 
an O $2p$ electron comes from one of the nearest neighbor oxygen sites 
to the Mn site and screens the potential due to the Mn $1s$ core hole. 
This screening process is confirmed by 
the theoretical calculation in the small cluster 
and the x-ray absorption spectroscopy experiments. \cite{ishi4,tranq,sheiner} 
Due to the hybridization between the O $2p$ and Mn $3d$ orbitals, 
this state strongly mixes with the state 
where an $e_g$ orbital in 
one of the nearest neighbor Mn sites ($j$ site) is empty.  
When the electron in the $4p$ orbital is relaxed to the $1s$ orbital with emitting x-ray, 
one of the $3d$ electrons in $i$ site comes back to $j$ site 
and the orbital excitation is brought about. 
This is represented by 
\begin{eqnarray}
| 3d_{i \gamma_i}^1 3d_{j \gamma_j}^1\rangle +h\nu  &\rightarrow& 
| 3d_{i  \gamma_i}^1 3d_{i \gamma_i'}^1   
\underline{1s_i} 4p_i^1   \rangle \nonumber \\
&\rightarrow& 
| 3d_{i \gamma_i''}^1  3d_{j \gamma_j'}^1  \rangle +h\nu' \ , 
\label{eq:kprocess}
\end {eqnarray}
where O $2p$ states is integrated out.  
Although this excitation process is similar to 
that of the Raman scattering due to the orbital wave, \cite{inou}  
the current operators are different between the two. 
Furthermore, 
in the Raman scattering for the conventional light, 
the polarization of the light determines the direction between 
$i$ and $j$ sites, unlike in the present case.  
In the case of manganites, the charge transfer from the nearest neighboring 
Mn sites occurs more easily than cuprates and nickelates, 
since the intra-site Coulomb interaction $(U)$ 
is smaller and the energy gain 
of the Hund coupling in the intermediate states 
($3d^{5}$ states) is larger in the present case. \cite{arima}
In the L-edge case the one-orbital wave excitation is only brought about. 
On the other hand, in the K-edge case, 
both one- and two-orbital wave excitations occur. 
\par
\subsection{\it L-edge case} 
The transition probability of RIXS for the orbital wave 
is calculated in the electron-photon coupled system with the 
Hamiltonian, 
\begin{equation}
H=H_{ele}+H_{ph}+H_{e \mbox{-} p} \ . 
\label{eq:hamil}
\end{equation}
The first term is the electronic one for  
the Mn $3d$ and Mn $2p$ orbitals which 
includes the following three terms: 
\begin{equation}
H_{ele}=H_{3d}+H_{2p}+H_{3d \mbox{-} 2p} \ , 
\label{eq:hele}
\end{equation}
with 
\begin{eqnarray}
H_{3d}&=& \varepsilon_d \sum_{i \gamma_\theta \sigma} 
d^\dagger_{i \gamma_\theta \sigma} d_{ i \gamma_\theta \sigma}
+\sum_{\langle ij \rangle \gamma \gamma' \sigma} t_{ij}^{\gamma \gamma'}
d_{i \gamma \sigma}^\dagger d_{j \gamma' \sigma} 
\nonumber \\
&+&U\sum_{i \gamma_\theta} \ 
    n_i(3d_{ \gamma_\theta \uparrow}) \  n_i(3d_{\gamma_\theta \downarrow})  \nonumber \\
&+&U' \sum_{i \sigma \sigma' \gamma_\theta} 
n_i(3d_{\gamma_{\theta+ } \sigma }) \  n_i(3d_{\gamma_{\theta- } \sigma'}) \nonumber \\
&+&J\sum_{i \sigma \sigma' \gamma_\theta}   
d_{i \gamma_{\theta+}  \sigma }^\dagger   d_{i \gamma_{\theta-}  \sigma'}^\dagger
d_{i \gamma_{\theta+}  \sigma'}           d_{i \gamma_{\theta-}  \sigma}  
\nonumber \\
&+& H_{e \mbox{-} t}+H_{t \mbox{-} t} \ , 
\label{eq:h3d}
\end{eqnarray}
and 
\begin{eqnarray}
H_{2p}+H_{3d \mbox{-} 2p}&=& \varepsilon_p \sum_{i \gamma \sigma} 
p^\dagger_{i \gamma \sigma} p_{i \gamma \sigma}  
\nonumber \\
&+&\sum_{i \gamma_\theta \gamma } V(3d_{\gamma_\theta},2p_{\gamma}) 
n_i(3d_{\gamma_\theta}) n_{hi}(2p_{ \gamma}) \ .  
\label{eq:h3d2p}
\end{eqnarray}
$n_i(3d_{\gamma \sigma})$ and 
$n_{hi}(2p_{\gamma \sigma})$ are the number operators 
for Mn $3d$ electron and Mn $2p$ hole, respectively. 
$V(3d_{\gamma_\theta},2p_{\gamma}) $ 
is the Coulomb interaction between 
Mn $3d$ and $2p$ and is given by 
$ V(3d_{\gamma_{\theta }},2p_\gamma)= 
F_0(3d,2p)  -
4F_2 (3d,2p)\cos \Bigl(\theta+n_\gamma {2 \pi \over 3} \Bigr) $. \cite{ishi4}
The second and third terms in Eq. (\ref{eq:hamil}) describe 
the photon part and electron-photon interaction, respectively, 
and a sum of them is given by 
\begin{eqnarray}
H_{ph}&+&H_{e \mbox{-} p}=\sum_{k \lambda} \omega_{k} 
\Bigl( a^\dagger_{k \lambda} a_{k \lambda} +{1 \over 2} \Bigr) \ , 
\nonumber  \\
&-&\sum_{k \lambda} \sqrt{{2\pi \over \Omega }}  {1 \over \sqrt{\omega_k} }
 \vec j_k \cdot \vec e_{ k \lambda} 
 \Bigl(e^{-i\omega_{k} t} a_{ k \lambda}
+e^{ i\omega_{k} t} a_{ k \lambda}^\dagger \Bigr) \ ,  
\label{eq:heleph}
\end{eqnarray}
where 
$a_{ k \lambda}^\dagger$ is 
the photon creation operator with 
momentum $\vec k$ and polarization $\lambda$. 
In the L-edge case, 
the current operator describes  
the $2p \rightarrow 3d $ transition as  
\begin{equation}
(\vec j_{k})_{\alpha}=
\sum_{i \gamma_\theta \sigma} A^{(L)}_{i \gamma_\theta \alpha}
d_{i \gamma_\theta \sigma}^\dagger p_{i \alpha \sigma} 
e^{i \vec k \cdot \vec R _i}+ H.c. \ . 
\label{eq:jl}
\end{equation}
The coupling constant depends on the 
polarization and the orbital as follows,  
\begin{equation}
A^{(L)}_{i \gamma_\theta \alpha}=
A_0 \cos \Bigl ({\theta \over 2}+n_\alpha{2 \pi \over 3} \Bigr ) .  
\label{eq:aaa}
\end{equation}
with $(n_x,n_y,n_z)=(1,2,3)$.  
$A_0$ is a constant part defined by 
\begin{equation}
A_0=i{e \over m} \Biggl( {3 \sqrt{5} \over 4 \pi} \Biggr) 
\biggl \langle {-6 z^2x^2 \over r^5} 
+ { x^2z^2 \over r^4 } \nabla_r 
- { x^4 \over r^4} \nabla_r \biggr \rangle
\label{eq:a0}
\end{equation}
where $\langle A \rangle=\int dr R_{2p}(r) A R_{3d}(r)$.  
\par
Being based on the Hamiltonian, 
the scattering matrix for the orbital wave is derived. 
The initial and final states 
are described by $\widetilde H_{3d}$ (Eq.(\ref{eq:hamiltonian})) 
where the orbital wave is the eigen-mode. 
The intermediate state is described by the single site term in $H_{ele}$ 
(Eq.(\ref{eq:hele})) and the inter-site effects are neglected. 
The explicit form of the inelastic part of the scattering matrix is 
obtained as 
\begin{eqnarray}
(S)_{fi}&=& -2\pi i \delta (E_f-E_i) \nonumber \\
&\times& {2 \pi \over \Omega} 
\sum_{k_1 k_2 \lambda_1 \lambda_2} {1 \over \sqrt{\omega_{k_1} \omega_{k_2}}}
\sum_{\alpha \  i}  
(e_{k_1 \lambda_1})_\alpha (e_{k_2 \lambda_2})_\alpha  \nonumber \\
&\times & e^{i(\omega_{k_1} -\omega_{k_2})t-i(\vec{k_1} -\vec{k_2}) \cdot \vec R_i}
I_{i \alpha}  D \ . 
\label{eq:sl}
\end{eqnarray}
$E_{i(f)}$ is the energy at the initial (final) state in the electron-photon 
system. 
In Appendix A, 
we discuss the elastic part which is a probe to detect the orbital ordering. 
In the above formula, 
the orbital excitation by x-ray 
is represented by the operator, 
\begin{equation}
I_{i \alpha }=
(A_{i a \alpha}^{(L) \ast} A_{i b \alpha}^{(L)} T_{i-} + 
A_{i b \alpha}^{(L) \ast} A_{i a \alpha}^{(L)} T_{i+}) \ , 
\label{eq:il}
\end{equation}
where subscript $a(b)$ indicates the occupied (unoccupied) $3d$ orbital 
at site $i$. 
$D$ is the energy denominator in the perturbation given by 
\begin{equation}
D=-{3 \over 2} E^{(t) -1}+{1 \over 2} E^{(s) -1} \ . 
\label{eq:csigma}
\end{equation}
Two terms correspond to 
the different intermediate states where  
two electrons occupy the different $e_g$ orbital with the 
triplet and singlet spin states, respectively (see Appendix A). 
It is mentioned that 
when the wave function is chosen to be real, 
$I_{i \alpha}$ in Eq.(\ref{eq:il}) is expressed by $T_x$, 
and it is in contrast to the neutron scattering where 
both $S_x$  and $S_y$ appear in the scattering matrix. 
This remarkable feature brings about 
the selection rule between the reciprocal lattice vector and 
the mode of the orbital wave as we will show later. 
The transition probability is 
obtained by the conventional golden rule. 
By rewriting the pseudo-spin operator 
in Eq.(\ref{eq:il}) by the Holstein-Primakoff bosons, 
we obtain the transition probability 
in the L-edge case: 
\begin{eqnarray}
W&=& C \sum_{G q \mu } |F_{\mu \alpha} (\vec q, \vec G)|^2 \nonumber \\
&\times& 
\bigl ( \delta_{\vec k_1-\vec k_2+\vec q+\vec G}
        \delta(\omega+\omega_q^{(\mu)}) (1+n_q^{(\mu)})   \nonumber \\
       &+& \delta_{\vec k_1-\vec k_2-\vec q+\vec G} 
        \delta(\omega-\omega_q^{(\mu)} )  n_q^{(\mu)} \bigr )   \ ,  
\label{eq:wl2}
\end{eqnarray}
with $C=(2\pi / \Omega \sqrt{\omega_{k_1} \omega_{k_2}} )^2$. 
$n_q^{(\mu)}$
is the number of the orbital wave of  
mode $\mu$ and $\vec G$ is the reciprocal lattice vector. 
Since the one orbital wave process is 
possible in the present case, 
the transition probability shows a similar 
form to that for the one magnon neutron scattering. 
Difference in the present case from the magnon scattering 
reflects on the generalized structure factor: 
\begin{equation}
F_{\mu \alpha}(\vec q, \vec G)=D \sum_{\nu} A_{\nu a \alpha}^{(L) \ast} 
A_{\nu b \alpha}^{(L)}
(V_{\nu \mu}(\vec q)+W_{\nu \mu}(\vec q) ) e^{i \vec G \cdot \vec r_\nu} \ , 
\label{eq:stl}
\end{equation}
where $V_{\nu \mu}(\vec q)$ and $W_{\nu \mu}(\vec q)$ are the coefficients 
in the Bogolyubov transformation which connect the operator 
for the $\nu$-th ion to that for the $\mu$-th eigen mode as 
\begin{equation}
a_{\nu q}=\sum_{mu} V_{\nu \mu}(\vec q) \alpha_{\mu q}
+W_{\nu \mu}(\vec q) \alpha_{\mu -q}^\dagger \ . 
\label{eq:vw}
\end{equation}
It is noticeable that in the above formula, 
$\vec G$-, $\vec q$-, and $\alpha$-dependence of the structure 
factor is mainly dominated by the factors $e^{i \vec G \cdot \vec r_\nu}$, 
$V_{\nu \mu}(\vec q)+W_{\nu \mu}(\vec q)$ 
and $A_{\nu \gamma \alpha}^{(L)}$, respectively.  
\par
In Fig. 5, numerical results of the structure factor for 
the spin-F case is presented. 
The orbital state is chosen as $(\theta_A/\theta_B)=({\pi \over 2}  /  {\pi \over 2}+\pi)$. 
$\vec G=(h,k,l)$ for the fcc lattice 
is chosen as $h+k+l=\mbox{even}$ (Fig. 5(a)) and 
odd (Fig. 5(b)), respectively. 
It is clear that the $(+)$- and $(-)$-modes are observed separately 
in the odd and even cases,  
and their weights are proportional to 
$(\cosh \theta^{(+)}_q+\sinh \theta^{(+)}_q)^2$ and 
$(\cosh \theta^{(-)}_q+\sinh \theta^{(-)}_q)^2$, respectively. 
This selection rule is originated from 
the following facts: 1) 
when the wave functions of the $e_g$ orbitals 
are chosen to be real, the scattering matrix in Eq.(\ref{eq:sl}) and (\ref{eq:il})
is represented by $T_x$, 2) the absolute value of the coupling constants 
$|A^{(L)}_{\nu a \alpha} A^{(L)}_{\nu b \alpha}|^2$
in the two sublattices is 
the same in the orbital order $(\theta_A/\theta_A+\pi)$, 
3) the relative motion of the pseudo-spin in $(+)$- 
and $(-)$-modes is opposite, 
and 4) there exists the factor 
$e^{i \vec G \cdot \vec r_\nu}$.  
It is shown that 
this selection rule does not depend on the 
phase of the wave functions of the $e_g$ orbitals. 
It is highly in contrast to the case of one magnon neutron scattering 
where the correlation function of $S_y$ breaks the selection rule 
and the $(+)$- and $(-)$-modes are also 
observed in the $h+k+l=\mbox{even}$ and odd 
cases with weights of 
$(\cosh \theta^{(+)}_q-\sinh \theta^{(+)}_q)^2$ and  
$(\cosh \theta^{(-)}_q-\sinh \theta^{(-)}_q)^2$, respectively.
The intensity of the structure factor monotonically 
increases with decreasing the energy and it diverges 
at the $\Gamma$-point. 
This $\vec q$-dependence comes from 
$V_{\nu \mu}(\vec q)+W_{\nu \mu}(\vec q)$ 
which is proportional to 
$(\cosh \theta_{q}^{(\mu)}+\sinh \theta_{q}^{(\mu)})^2$. 
The divergence at the $\Gamma$-point is originated from 
the divergence of the $x$ component of the staggered orbital 
order parameter  
due to the rotational symmetry 
in the $(T_z,T_x)$-plane.  
In the figure, 
the intensity in the $z$-polarization 
is larger than that in the 
$x$-polarization and the ratio does not depend on $\vec q$.
It is interpreted that 
the orbital order 
is close to $(3d_{3x^2-r^2}/3d_{y^2-z^2})$  where 
$A^{(L)}_{y^2-z^2 \ x}$ is zero due to the symmetry. 
In Fig. 6, the structure factor in the spin-A 
case with the $x$-polarization is shown. 
The four modes of the orbital wave, 
that is, $(+ \pm)$- and $(- \pm )$-modes  
are mainly observed for $h+k$=odd, $l$=odd (even), and 
$h+k$=even, $l$=odd (even), respectively. 
Strictly speaking, 
this selection rule is satisfied in the case of $J_2=0$ and 
in general, (++)- and (- -)-modes and 
(+-)- and (-+)-modes are mixed each other. 
The intensity does not diverge  
at the $\Gamma$-point, unlike that in the spin-F case, 
since the rotational symmetry in the orbital space is broken by the anisotropic spin 
structure. 
The intensity is larger in the presented $z$-polarization case  
in comparison with that in the $x$-polarization one, 
although the results in the $x$-polarization are only shown in the figure. 
It is attributed to the type of the orbital order  
which is close to $(3d_{3x^2-r^2}/3d_{y^2-z^2})$ 
rather than $(3d_{3z^2-r^2}/3d_{x^2-y^2})$. 
\subsection{\it K-edge case}
In the K-edge case, 
we derive 
the effective Hamiltonian for the electron-photon 
system and calculate the transition probability by using the 
Hamiltonian.  
The electronic part of the Hamiltonian, 
where Mn $3d$, $4p$ and $1s$ orbitals are introduced, 
is given by \cite{ishi4}
\begin{eqnarray}
H_{ele}=H_{0}+H_{3d}+H_{3d \mbox{-} 4p}+H_{1s \mbox{-} 3d} +H_{1s \mbox{-} 4p}  \ , 
\label{eq:helek}
\end{eqnarray}
with 
\begin{eqnarray}
H_{0}=
\varepsilon_P  
\sum_{i \gamma \sigma} 
P^\dagger_{i \gamma \sigma} P_{i \gamma \sigma} 
+ \varepsilon_s  
\sum_{i \sigma} 
s^\dagger_{i \sigma} s_{i \sigma}  \ , 
\label{eq:h1k}
\end{eqnarray}
where $s_{i \sigma}$ and $P_{i \gamma \sigma}$ are 
the annihilation operator of Mn $1s$ and $4p$ electron 
with spin $\sigma$ and orbital $\gamma$, respectively. 
The last three terms in Eq.(\ref{eq:helek})
describe the Coulomb interactions: 
\begin{eqnarray}
H_{3d \mbox{-} 4p}&+&H_{1s \mbox{-} 3d}+H_{1s \mbox{-} 4p} \nonumber \\
&=&\sum_{i \gamma_\theta \gamma} V(3d_{\gamma_\theta},4p_{\gamma}) 
n_i(3d_{\gamma_\theta}) n_i(4p_{\gamma})   \nonumber \\
&+&\sum_{ i \gamma_\theta } V(3d,1s) 
n_i(3d_{\gamma_\theta}) n_{hi}(1s)   \nonumber \\
&+&\sum_{ i \gamma } V(4p,1s) 
n_i(4p_{\gamma}) n_{hi}(1s)  \ . 
\label{eq:hab}
\end{eqnarray}
In the present case, 
the current operator in $H_{e \mbox{-} p}$ describes the 
$1s \rightarrow 4p$ transition as \cite{ishi4}
\begin{equation}
(\vec j_{k})_{\alpha}=
\sum_{i \sigma} A^{(K)}
P_{i \alpha \sigma}^\dagger s_{i \sigma} 
e^{i \vec k \cdot \vec R _i}+ H.c. \ . 
\label{eq:jk}
\end{equation}
$A^{(K)}$ is the coupling constant defined by 
\begin{equation}
A^{(K)}={e \over m} \int d \vec r \ \phi_{4p_\alpha}^\ast (\vec r) (-i \nabla_\alpha)
\phi_{1s}(\vec r) \ , 
\label{eq:asp}
\end{equation}
which is derived by the dipole approximation 
and does not depend on the polarization $\alpha$. 
\par
The scattering process in the K-edge case 
includes the electron transfer as shown in Fig. 4(b) 
and is derived by the fourth order process with respect to 
$H_{e \mbox{-} p}$ and $H_t$ (the transfer term in $H_{3d}$ (Eq.(\ref{eq:h3d})). 
Therefore, in order to avoid complications 
due to the higher order perturbation, 
we derive the effective Hamiltonian by the following procedure. 
In the initial and final states of the scattering, 
the energy of the doubly occupied $e_g$ state is higher 
than that of the singly occupied one by the order of 
the Coulomb interaction between $3d$ electrons. 
In the intermediate state, on the other hand, 
due to the screening effects, 
the energy of the singly occupied state is higher 
that that of the doubly occupied one by 
the order of 
(the screening potential)-(the Coulomb interactions 
between $3d$ electrons and $3d$ and $4p$ electrons). 
By excluding these two states which have the higher energy eigen values,  
we derive the effective Hamiltonian by the second order perturbational 
calculation of $H_{t}$ and/or $H_{e \mbox{-} p}$ 
as follows, 
\begin{equation}
\widetilde H=H_{e \mbox{-} e}+ H_{core}+\widetilde H_{e \mbox{-} p}+H_{ph}. 
\label{eq:heff}
\end{equation}
The detailed derivation is presented in Appendix B. 
The first term was introduced in Eq.(\ref{eq:effect}) 
and the second term describes the electronic states 
where the $1s$ core hole and the $4p$ electron always exist. 
The third term is recognized as the effective electron-photon interaction 
given by 
\begin{eqnarray}
{\widetilde H_{e \mbox{-} p}}&=&- \sqrt{2 \pi \over \Omega} 
\sum_{i \delta} \sum_{m  k \lambda \alpha} 
\sum_{\gamma \gamma' \gamma'' \sigma \sigma'} 
{1 \over \sqrt{\omega_k}}
t_{ii+\delta}^{\gamma \gamma''} \Gamma^{(m)}_{\gamma \alpha} \nonumber \\
&\times& \bigl (P_m d^\dagger_{i \gamma' \sigma'} n_i(3d_{\gamma \sigma})
d_{i+\delta \gamma'' \sigma'}  \bigr ) \nonumber \\
&\times& j_{i \alpha k}   a_{k \lambda} (e_{k \lambda})_\alpha 
e^{-i \omega_k t}
+ H.c. \ .  
\label{eq:heffep}
\end{eqnarray}
It implies the electron excitation from $1s$ to $4p$ 
accompanied with the electron transfer from 
$i+\delta$ site to $i$ site, 
where the operator $P_m$ projects out the $m$-th doubly occupied state. 
$\Gamma^{(m)}_{ \gamma \alpha}$ is the energy denominator 
derived in the perturbational calculation (Eq.(\ref{eq:gamma})) 
and $( j_{i k})_\alpha$ is the $i$-th component in Eq.(\ref{eq:jk}). 
\par
The scattering matrix 
is calculated by the perturbation  
of the effective interaction ${\widetilde H_{e \mbox{-} p}}$  
and is given by 
\begin{eqnarray}
(S)_{fi}&=&
-2 \pi i \delta(E_f-E_i)
\\ \nonumber 
& \times& 
{2 \pi \over \Omega} \sum_{k_1 k_2 \lambda_1 \lambda_2 \alpha} 
{1 \over \sqrt{\omega_{k_1} \omega_{k_2}}} \sum_{ i \sigma}
(e_{k_1 \lambda_1})_\alpha (e_{k_2 \lambda_2})_\alpha  \nonumber \\
&\times& e^ {i (\omega_{k_1}-\omega_{k_2}) t - i(\vec k_1-\vec k_2) \cdot \vec R_i} 
 \nonumber \\
& \times & |A^{(K)}|^2 \sum_\delta^{'} \sum_m  I^{(m)}_{i \delta \alpha}  D^{(m)}_{\alpha}  \ . 
\label{eq:sk}
\end{eqnarray}
$D_{\alpha}^{(m)}$ and $I^{(m)}_{i \delta \alpha}$ are 
defined in Appendix B. 
$I^{(m)}_{i \delta \alpha}$ is the main part of the formula 
and is expressed by 
the spin and orbital pseudo-spin operators in $i$ and its 
nearest neighbor $i+\delta$ sites. 
It is noticeable that $I^{(m)}_{i \delta \alpha}$ is described by 
$T_x$ and $T_z$ (see Appendix B), 
as well as that in the L-edge case in Eq.(\ref{eq:il}). 
It describes changes of the 
spin and orbital states in these sites.  
The scattering processes, where the one and two orbital waves 
are concerned, 
are included in $I^{(m)}_{i \delta \alpha}$ and these processes occur in the same order. 
This is in contrast to the conventional magnon Raman scattering 
where the two magnon sector is dominant. 
This is attributed the fact that $t_{ij}^{\gamma \gamma'}\ne \delta_{\gamma \gamma'}$. 
Therefore, we consider the processes for the one orbital wave excitation 
in $I^{(m)}_{i \delta \alpha}$ schematically shown in Fig. 7. 
In this approximation, 
the transition probability is obtained as 
the same form as Eq. (\ref{eq:wl2}) 
with the structure factor replaced by 
\begin{eqnarray}
F_{\mu \alpha}(\vec q, \vec G)&=&2\sum_{\nu m} D_{\alpha}^{(m)} e^{i \vec G \cdot \vec r_\nu} |A^{(K)}|^2
\nonumber \\ 
&\times&
\sum_{\nu'} 
\Bigl( 
\bigl( V_{\nu \mu}(\vec q)+W_{\nu \mu}(\vec q) \bigr) C_{\nu \nu'}^{(m)}
\nonumber \\
&+&e^{i \vec q \cdot (\vec r_\nu-\vec r_{\nu'} ) }
\bigl( V_{\nu' \mu}(\vec q)+W_{\nu' \mu}(\vec q) \bigr) 
B_{\nu \nu'}^{(m)}    \Bigr ) \ .  
\label{eq:fk}
\end{eqnarray}
The index $\nu'$ indicates one of the 
nearest neighboring Mn ions of the $\nu$-th ion 
and $B_{\nu \nu'}^{(m)}$ and $C_{\nu \nu'}^{(m)}$ 
are represented by the spin correlation function 
defined in Appendix B.
The two terms in the parenthesis in Eq. (\ref{eq:fk}) corresponds to the 
orbital excitation in $\nu$ site and its nearest neighboring $\nu'$ site, 
respectively. 
As a result, 
the factor $e^{i \vec q \cdot (\vec r_\nu- \vec r_{\nu'} ) }$ 
gives the additional $\vec q$-dependence in the structure factor. 
\par
In Fig. 8, the numerical results of the structure factor in the spin-F case 
are presented. 
The parameter values are chosen as $w_0/t_0=0.5$, $W/t_0=-6$, 
$\tilde U/t_0=5$, $J/t_0=1.0$, and $\Gamma/t_0=0.5$ (see Appendix B), although 
the qualitative results do not depend on the parameter values. 
$h+k+l$=odd and even cases 
mainly correspond to 
$(+)$- and $(-)$-modes, respectively, 
although the both intensities are strongly mixed each other. 
The $\vec q$-dependence of the intensity 
is not monotonic, unlike that in the L-edge 
case shown in Fig. 5. 
Especially, around the $\Gamma$-point, 
the intensity gradually decreases and 
becomes zero at the point. 
This is attributed to the following unique excitation process 
in the K-edge case.   
With using the 
relations $\sum_{l} t^{aa}_{i i+l} t^{ba}_{i i+l}=0$ 
and $\sum_{l} t^{bb}_{i i+l} t^{ba}_{i i+l}=0$, 
in the uniform spin structure, 
$\sum_{\nu'} C_{\nu \nu'}=0$ 
and $\sum_{\nu'} B_{\nu \nu'}=0$ 
are derived. 
These relations suppress the divergence of the factor 
$V_{\mu \nu}(\vec q)+W_{\mu \nu}(\vec q)$ 
at the $\Gamma$-point in Eq. (\ref{eq:fk}), 
so that the intensity becomes zero at the point. 
It implies that at the $\Gamma$-point 
the several orbital excitation processes 
by x-ray are canceled out 
each other. 
The other characteristic $\vec q$-dependences of the 
intensity, such as, 
no intensity along the $\Gamma-L$ direction, 
are interpreted by the factor 
$e^{i \vec q \cdot (\vec r_\nu-\vec r_{\nu'} ) }$. 
The polarization dependence 
of the intensity comes 
from the Coulomb interaction 
$V(3d_{\gamma}, 4p_\alpha)$ 
included in $\Gamma^{(m)}_{\gamma \alpha}$, 
although it is not remarkable in comparison with that 
in the L-edge case (Fig. 9). 
Around the $X$-point, 
the slightly larger intensity 
is shown in the $x$-polarized case 
in comparison with that in the $z$-polarized one 
and it is interpreted as follows. 
In the present orbital order, 
a dominant excitation process around $X$-point is 
that in Fig. 5(b), where 
the orbital states in $i$ and $j$ sites are $\theta={\pi \over 2}+\pi$ 
and ${\pi \over 2}$, respectively. 
Since the orbital at the $i$-site, that is, 
$ {1 \over \sqrt{2}} 
(3d_{3z^2-r^2}+3d_{x^2-y^2}) $, 
shrinks in $x$-direction, 
the $1s \rightarrow 4p_x$ excitation 
shows a larger intensity than that of $1s \rightarrow 4p_{y(z)}$ 
excitation due to the Coulomb interaction. 
Therefore, the intensity is slightly larger in the $x$-polarized case. 
In Fig. 10, the results in the spin-A case 
with the $x$-polarization are presented. 
The relation between the reciprocal lattice vector and the four excitation modes 
is the same as that in the L-edge case. 
However, the two modes 
are mixed and the mixing becomes remarkable around 
the $M$-point due to the factor 
$e^{i \vec q \cdot (\vec r_\nu- \vec r_{\nu'} ) }$. 
The intensity of the structure factor 
does not show a strong polarization dependence of 
x-ray in comparison with that in the L-edge case, 
since the polarization dependence is 
originated from that in $\Gamma^{(m)}_{\gamma \alpha}$ 
as explained above. 
\par
\section{summary and discussion}
In this paper, we study the orbital excitation in the orbital ordered manganites 
and the method to observe the excitation by using the resonant inelastic 
x-ray scattering. 
At first, we emphasize the noticeable correlation between the static spin structure and the 
orbital excitation. 
The anisotropic spin ordering brings 
about the anisotropy in the orbital space and causes the gap in the orbital excitation. 
In this case, the interaction between spin wave and phonon 
seems to be weak, since the gap in the orbital 
excitation is of the order of $t_0^2/U$. 
This situation is expected to be realized in $\rm LaMnO_3$ where 
the orbital ordered temperature accompanied with the structural phase 
transition occurs $T_O=780K$ and the A-type antiferromagnetism is 
realized at $T_N=140K$. \cite{mura2}
Although the coupling with the lattice contributes to the gap, 
it is predicted that the nature of the orbital excitation 
is changed at $T_N$ and below the 
temperature the gap becomes more remarkable. 
On the other hand, in the spin-F case, 
the orbital excitation becomes gapless and 
the low lying excitation affects the thermodynamic 
and transport
properties. 
In this case, 
the dispersion relation is modified by 
the interaction between spin wave and phonon. 
The specific heat which is proportional to $T^3$ 
is derived by the linear dispersion of the 
orbital wave, 
which is distinguished from the contribution 
from the ferromagnetic spin wave. 
The possible candidate, where the gapless excitation 
is anticipated, 
is the ferromganetic 
insulating phase realized in the low hole doped manganites,  
where the superexchange interaction between $e_g$ orbitals 
dominates the ferromagnetic interaction 
rather than the double exchange one. \cite{maezo,mura3}
Especially, in $\rm La_{0.88}Sr_{0.12}MnO_3$, 
the orbital ordering is confirmed by the resonant elastic x-ray scattering 
below $T_L=140K$, 
and the static Jahn-Teller distortion is released in this 
temperature range. \cite{mura3}
Therefore, the low lying excitation is expected to be observed without  
disturbance of the lattice distortion. 
In the spin canted phase among the above two concentration region, 
the anisotropy of the spin correlation function decreases and 
the gap in the orbital excitation gradually decreases with 
doping holes. 
Such kind of control of the orbital excitation is possible 
by changing the 
spin structure in the magnetic field. 
\par
As a probe to detect the orbital excitation, 
we propose the two possible methods by using the 
inelastic resonant x-ray scattering. 
In the present case, the dispersion relation is detectable due to the smaller 
wave length of the x-ray in comparison with the conventional light 
(the wave lengths in the L- and K-edge cases are about 2.1$\AA$ and 19.6$\AA$, 
respectively). 
In the L-edge case, 
the orbital excitation is directly caused by $2p \rightarrow 3d$ transition 
and the single orbital wave without the spin flipping is only brought about. 
In the spin-F case with orbital $(\theta/\theta+\pi)$, 
each mode of the orbital wave is separately observed 
by choosing the reciprocal lattice vector $\vec G$.  
This selection rule between the reciprocal lattice vector and 
the modes of the orbital wave is mainly due to 
the fact that the inelastic part of the scattering matrix 
is represented by the $x$ component of the pseudo spin operator. 
Since the large scattering intensity is expected 
in the low energy region near the orbital superlattice reflection point, 
it may be observed without disturbance of the 
large fundamental elastic peak. 
The polarization dependence is remarkable, 
since it directly comes from the coupling constant 
between electron and photon. 
In the K-edge case, where the elastic scattering has already 
been used to observe the 
orbital ordering, 
the orbital excitation occurs through the 
screening process of the core hole. 
This process through the charge transfer from 
one of the nearest neighboring Mn ions
is more favorable in manganites rather than 
in cuprates and nickelates, 
since the manganites are closer to 
the boundary between the 
charge transfer insulator and the Mott one. 
The one and two orbital wave excitations are possible 
in the same order, unlike the conventional two magnon 
Raman scattering. 
The orbital excitation occurs in the site 
where the x-ray excites the $1s$ electron 
and its one of the nearest neighboring Mn sites. 
Owing to the unique excitation process, 
the structure factor shows 
the strong momentum dependence 
and in particular the intensity vanishes around the $\Gamma$-point. 
Through the measurement of the characteristic momentum dependence, 
it is possible to identify the orbital sector in the inelastic component 
and the mechanism of the orbital excitation proposed in this paper. 
\par
\medskip
\noindent
ACKNOLEDGEMENTS
\par
The authors would like to thank Y. Endoh, Y. Murakami 
and J. Mizuki for their valuable 
discussions. 
We also indebted to M. Kaji and S. Okamoto for their helpful 
discussions and calculations. 
This work was supported by Priority Areas Grants from the Ministry of 
Education, Science and Culture of Japan, CREST (Core Research for Evolutional 
Science and Technology Corporation) Japan, and NEDO Japan. Part of the numerical 
calculation was performed in the HITACS-3800/380 supercomputing 
facilities in Institute for Materials Research, Tohoku University. 
\appendix
\section{elastic scattering in the L-edge case}
In this appendix, we introduce the elastic x-ray scattering 
in the L-edge case  
as a method to detect the orbital order. 
We focus on the 
polarization and orbital dependence of the atomic scattering factor 
in the orbital ordered phase. 
Once the dependence is revealed, the scattering intensity in the several orbital 
ordering is obtained by the formulas in Ref. \onlinecite{ishi5}. 
The atomic scattering factor is calculated by the perturbational calculation 
presented in Sect. IIIA and given by  
\begin{eqnarray}
f_{i \alpha \alpha}
&=&|A_{\gamma_{\theta -} \alpha}^{(L)}|^2 {3 \over 2}  E^{(t)-1} 
\nonumber \\
&+&|A_{\gamma_{\theta -} \alpha}^{(L)}|^2  {1 \over 2}  E^{(s)-1}
+|A_{\gamma_{\theta} \alpha}^{(L)}|^2 E^{(d)-1} \ .  
\label{eq:fela}
\end{eqnarray}
As well as in the inelastic case, 
we approximately set up the state and the energy 
for the intermediate states in the scattering. 
We assume that the states consist of 
the following three states: 
$|t \underline{2p_\alpha} \rangle$, 
$|s \underline{2p_\alpha} \rangle$, and 
$|d_\gamma \underline{2p_\alpha} \rangle$, 
where $| s \rangle (| t \rangle )$ represents the state where 
two $e_g$ electrons 
occupy the different orbital with the spin singlet (triplet) 
and $ | d_\gamma \rangle$ is 
the doubly occupied state in the $\gamma$ orbital.
These energies are approximately estimated as 
$E^{(t)}=U'-J+\Delta+2F_0(3d,2p)-\omega_k+i\Gamma$ with 
$\Delta=\varepsilon_d-\varepsilon_p$, 
$E^{(s)}=U'+J+2J_H+\Delta+2F_0(3d,2p)-\omega_k+i \Gamma $ 
and 
$E^{(d)}=U+2J_H+\Delta+2V(3d_\gamma,2p_\alpha)-\omega_k+i\Gamma $, 
where $\Gamma$ is the damping constant. 
Because $E^{(t)}$ is the smallest, 
$|t \underline{2p_\alpha} \rangle$ dominates the L-edge. 
It is noticed that the energy $E^{(t)}$ 
is independent of 
the occupied $3d$ orbital and the polarization of x-ray, because 
two $e_g$ orbitals are occupied in the intermediate state 
and three $2p$ orbitals are occupied in the initial and final states. 
On the other hand, its intensity in Eq.(\ref{eq:fela}) is proportional to 
$|A_{\gamma_{\theta -} \alpha }^{(L)}|^2$ which 
clearly depends on the occupied orbital and the polarization as 
shown in Eq.(\ref{eq:aaa}). 
It is concluded that 
the anisotropy of the scattering factor 
is originated from the 
coupling constant between electron and photon, 
unlike that in the K-edge case 
where the Coulomb interaction between $3d$ and $4p$ gives 
rise to the anisotropy. \cite{ishi4}
For example, 
In the antiferro-type orbital ordering, where 
two kinds of the orbital sublattice exist, 
the scattering intensity at the orbital superlattice reflection 
is proportional to 
$|(|A_{\gamma_{A-} \alpha}|^2-|A_{ \gamma_{B-} \alpha}|^2 )|^2$, 
where $\gamma_{A(B)-}$ is the unoccupied 
orbital state in the A(B)-orbital sublattice. 
\section{detailed derivation and formulas in K-edge case}
In this appendix, we present the 
detailed derivation and formulation 
of the effective Hamiltonian 
and transition probability in the 
K-edge case which are introduced in Sect. III B. 
\par
In derivation of the effective Hamiltonian, 
the unperturbed states are classified 
by the number of electron of Mn $3d$ and $4p$ orbitals at each site 
as $| n(3d) , n(4p) \rangle$ 
and these states and energies are approximately set up 
as follows.  
The energy in  $| 1,0 \rangle$ is defined as the origin of the energy. 
$| 1,1 \rangle $ consists of 
$| 3d_{\gamma}^1 4p_{\gamma' }^1 \underline{1s}  \rangle$ 
and its energy is difined as  
$E_{11}=\Delta+V(3d_{\gamma}, 4p_{\alpha})
+V(3d,1s)+V(4p,1s)-\omega_k $ 
with $\Delta=\varepsilon_{P}-\varepsilon_s$. 
$| 2,0 \rangle $ includes the three states represented by 
$| t \rangle$, $| s \rangle$ and $| d \rangle $, 
as we explained in Appendix A,  
with the energies 
$ E_{20}^{(t)}=U'-J $, 
$ E_{20}^{(s)}=U'+J+2J_H $ 
and 
$ E_{20}^{(d)}=U+2J_H $, 
respectively. 
In the same way, 
$| 2, 1 \rangle $ consists of the three states: 
$| t  \ 4p_\alpha^1 \underline{1s} \rangle $, 
$| s \ 4p_\alpha^1 \underline{1s} \rangle $ and 
$| d_\gamma \ 4p_\alpha^1 \underline{1s} \rangle $ 
with energies  
$ E_{21}^{(t)}=U'-J+\Delta+2F_0(3d,4p)
+V_{core}-\omega_k$, 
$ E_{21}^{(s)}=U'+J+2J_H+\Delta+2F_0(3d,4p)
+V_{core}-\omega_k$ 
and 
$ E_{21}^{(d)}=U+2J_H+\Delta+2V(3d_{\gamma},4p_\alpha)
+V_{core}-\omega_k$, 
respectively, 
with $V_{core}=2V(3d,1s)+V(4p,1s)$. 
In all states, 
the $t_{2g}$ spin is assumed to be parallel to the $e_g$ one and the 
spin dependence of the core hole interaction are neglected. 
Among these parameters, we choose  
the independent parameters $w_0=4F_2(4p,3d)$, $W'=U+2J+V(1s,3d)+F_0(4p,3d)-5J$
and $\tilde U=U-3J$ with 
$U'-J+\Delta+2F_0(3d,4p)+V_{core}-\omega_k=0$,  
$V(1s,3d)=V(1s,4p)$ and $J=J_H$. 
\par
The matrix element of the effective Hamiltonian $(\widetilde H)_{ln}$
is derived by the conventional canonical transformation. 
The perturbational processes are characterized by 
the set of the initial, intermediate and final states: 
$(| l \rangle , |m \rangle , | n \rangle)$.  
In the calculation, 
$| n(3d),n(4p) \rangle=\{ |1,0 \rangle, |2,1 \rangle \} $ are 
included explicitly and 
$\{ |2,0 \rangle, |1,1 \rangle \} $ are 
introduced in the virtual senses. 
$\widetilde H_{3d}$ in Eq.(\ref{eq:heff})
corresponds to the process 
$(| l \rangle , |m \rangle , | n \rangle)=
(|1,0 \rangle,|2,0 \rangle, |1,0 \rangle)$. 
The effective electron-photon interaction 
$\widetilde H_{e \mbox{-} p}$ in Eq.(\ref{eq:heffep}) is derived by 
$(|2,1 \rangle,|2,0 \rangle  |1,0 \rangle)$ and  
$(|2,1 \rangle,|1,1 \rangle, |1,0 \rangle)$. 
In this term, 
$\Gamma_{\gamma \alpha}^{(m)}$ is 
given by  
\begin{eqnarray}
\Gamma_{\gamma \alpha}^{(m)}={1 \over 2} \Bigl (
                         \bigl( E_{21}^{(m)}&-&E_{11} \bigr)^{-1}
                        +\bigl( E_{10}-E_{11} \bigr)^{-1}
\nonumber \\
                        +\bigl ( E_{21}^{(m)}&-&E_{20}^{(m)} \bigr)^{-1}
                        +\bigl( E_{10}-E_{20}^{(m)} \bigr)^{-1}  \Bigr ) \ . 
\label{eq:gamma}
\end{eqnarray}
The second term $H_{core}$ in Eq.(\ref{eq:heff})
is derived by the process 
$(|2,1 \rangle,|1,1 \rangle, |2,1 \rangle)$ 
and its 0-th order term of the perturbation 
is used in the calculation of the transition probability. 
The processes $(|1,0 \rangle,|1,1 \rangle, |1,0 \rangle)$ 
and $(|2,1 \rangle,|2,0 \rangle, |2,1 \rangle)$ 
imply the elastic off-resonance scattering 
and the higher order x-ray scattering, respectively, 
and these are irrelevant in the present scattering. 
\par
In the scattering matrix in Eq.(\ref{eq:sk}), 
$D_\alpha^{(m)}$ is the energy denominator 
originated from the second order 
perturbational process of $\widetilde H_{e \mbox{-} p}$
and it is represented by 
\begin{equation}
D_\alpha^{(m)}=\bigl( E_{21}^{(m)}-E_{10} +i\Gamma\bigr)^{-1} \ . 
\label{eq:ddd}
\end{equation}
$I_{ij}^{(m)}$ in Eq.(\ref{eq:sk}) describes the changes of the spin and 
orbital states defined as 
\par
\begin{eqnarray}
I^{(t)}_{i j \alpha}&=&\Bigl( {3 \over 4} + \vec S_i \cdot \vec S_j \Bigr)  
\nonumber \\
& \times &
\bigl(\Gamma^{(t)2}_{b \alpha} \tau_{i j}^A 
     +\Gamma^{(t)2}_{a \alpha} \tau_{i j}^B
-\Gamma^{(t)}_{a \alpha} \Gamma^{(t)}_{b \alpha} \tau_{i j}^C \bigr) \ , 
\end{eqnarray}
\begin{eqnarray}
I^{(s)}_{i j \alpha}&=&\Bigl({1 \over 4} - \vec S_i \cdot \vec S_j \Bigr)  
\nonumber \\
& \times &
\bigl(\Gamma^{(s)}_{b \alpha} \tau_{i j}^A 
    +\Gamma^{(s)2}_{a \alpha} \tau_{i j}^B
+\Gamma^{(s)}_{a \alpha} \Gamma^{(s)}_{b \alpha} \tau_{i j}^C \bigr) \ , 
\end{eqnarray}
and 
\begin{eqnarray}
I^{(d)}_{i j \alpha} &=& 2 \Bigl({1 \over 4} - \vec S_i \cdot \vec S_j \Bigr) 
\big( \Gamma^{(d)2}_{a \alpha}  \tau_{i j}^{A'} +\Gamma^{(d)2}_{b \alpha} 
 \tau_{i j}^{B'} \big) \ , 
\label{eq:III}
\end{eqnarray}
with 
\begin{equation}
\tau_{i j}^A=n_{ib} \bigl(t^{aa2}_{ij}n_{ja}+t^{ab2}_{ij}n_{jb}
+2t^{aa}_{ij} t^{ab}_{ij}T_{jx} \bigr) \ , 
\end{equation}
\begin{equation}
\tau_{i j}^B=n_{ia} \bigl(t^{ba2}_{ij}n_{ja}+t^{bb2}_{ij}n_{jb}
+2t^{bb}_{ij} t^{ba}_{ij}T_{jx} \bigr) \ , 
\end{equation}
\begin{equation}
\tau_{i j}^C=2 \bigl(t^{aa}_{ij}t^{ba}_{ij}T_{ix}n_{ja}+t^{bb}_{ij}t^{ab}_{ij}T_{ix}n_{jb}
+2t^{aa}_{ij}t^{bb}_{ij}T_{ix}T_{jx} \bigr) \ , 
\end{equation}
\begin{equation}
\tau_{i j}^{A'}=n_{ia} \bigl(t^{aa2}_{ij}n_{ja}+t^{ab2}_{ij}n_{jb}
+2t^{aa}_{ij}t^{ab}_{ij}T_{jx}
\bigr) \ , 
\end{equation}
\begin{equation}
\tau_{i j}^{B'}=n_{ib} \bigl(t^{ba2}_{ij}n_{ja}+t^{bb2}_{ij} n_{jb}
+2t^{bb}_{ij} t^{ba}_{ij}T_{jx} 
\bigr) \ . 
\end{equation}
\par
The coefficients $B_{\nu \delta}^{(m)}$ and $C_{\nu \delta}^{(m)}$ 
in Eq. (\ref{eq:fk}) 
are obtained by taking the average for the spin part as 
\begin{equation}
C_{\nu \nu'}^{(t)}=
-\Bigl( {3 \over 4}+ \langle \vec S_\nu 
\cdot \vec S_{\nu'}\rangle \Bigr)
\Gamma_{a \alpha}^{(t)} \Gamma_{b \alpha}^{(t)} t^{aa}_{\nu \nu'} t^{ba}_{\nu \nu'} \ , 
\end{equation}
\begin{equation}
C_{\nu \nu'}^{(s)}=
\Bigl( {1 \over 4}- \langle \vec S_\nu 
\cdot \vec S_{\nu'} \rangle \Bigr)
\Gamma_{a \alpha}^{(s)}\Gamma_{b \alpha}^{(s)} t^{aa}_{\nu \nu'} t^{ba}_{\nu \nu'} \ , 
\end{equation}
\begin{equation}
C_{\nu \nu'}^{(d)}=0  \ , 
\end{equation}
\begin{equation}
B_{\nu \nu'}^{(t)}=
\Bigl( {3 \over 4}+ \langle \vec S_\nu 
\cdot \vec S_{\nu'}\rangle \Bigr)
\Gamma_{a \alpha}^{(t)2} t^{bb}_{\nu \nu'} t^{ba}_{\nu \nu'} \ , 
\end{equation}
\begin{equation}
B_{\nu \nu'}^{(s)}=
\Bigl( {1 \over 4}- \langle \vec S_\nu 
\cdot \vec S_{\nu'} \rangle \Bigr)
\Gamma_{a \alpha}^{(s)2} t^{bb}_{\nu \nu'} t^{ba}_{\nu \nu'} \ , 
\end{equation}
and 
\begin{equation}
B_{\nu \nu'}^{(d)}=2
\Bigl( {1 \over 4}- \langle \vec S_\nu 
\cdot \vec S_{\nu'} \rangle \Bigr)
\Gamma_{a \alpha}^{(d)2} t^{aa}_{\nu \nu'} t^{ab}_{\nu \nu'} \ . 
\end{equation}
\par
\noindent
\vfill
\eject
\noindent
Figure captions
\par
\medskip \noindent
FIG. 1.  The mean field phase diagram at zero temperature. $F$, $A$, $C$, and $G$ imply the 
ferromagnetic structure, and the layer-type, rod-type and NaCl-type antiferromagnetic 
structures, respectively. 
\par
\medskip \noindent
FIG. 2. The dispersion relation of the orbital wave in the spin-F case. The Brillouin zone for the 
face-centered cubic lattice is adopted. 
The orbital states for the two orbital sublattices are 
denoted by $(\theta_A / \theta_A+\pi)$. 
\par
\medskip \noindent
FIG. 3. The dispersion relation of the orbital wave in the spin-A case. The Brillouin zone for the 
tetragonal lattice is adopted. $J_2/J_1=0.5$ which corresponds to the orbital state 
$(\theta_A/-\theta_A)$ with $\theta_A=1.34$. 
$\mu$ denotes the mode of the orbital wave. 
\par
\medskip \noindent
FIG. 4. The schematic picture of the 
orbital excitation processes in (a) the L-edge case and (b) the K-edge case. 
$j$ denotes one of the nearest neighbor sites of $i$. 
The broken arrows indicate the incident and scattered x-rays. 
\par
\medskip \noindent 
FIG. 5. The square of the structure factor for the orbital wave with spin-F 
in the L-edge case: 
$I_L=|F_{\mu \alpha}(\vec q, \vec G)|^2/(D|A_0|^2)^2$. 
(a) $h+k+l$=even, and (b) odd where $(h,k,l)$ is defined in 
a unit cell in the face-centered cubic lattice. The orbital 
state is chosen as $({\pi \over 2 }  /  {\pi \over 2}+\pi)$. The shaded circles and squares  
mean the $x$ and $z$ polarization cases, respectively. 
\par
\medskip \noindent
FIG. 6. The square of the structure factor for the orbital wave with spin-A in 
the L-edge case: 
$I_L=|F_{\mu \alpha}(\vec q, \vec G)|^2/(D|A_0|^2)^2$. 
(a) $h+k$=even and $l$=odd, and (b) $h+k$=odd and $l$=odd, 
where $(h,k,l)$ is defined in a unit cell in the tetragonal lattice. 
$J_2/J_1=0.5$ which corresponds to the orbital state 
$(\theta_A/-\theta_A)$ with $\theta_A=1.34$. 
The polarization is parallel to the $x$-direction.   
The open and filled circles mean (a) the $(+-)$- and $(-+)$-modes, 
and (b) the $(--)$- and $(++)$-modes, respectively. 
\par
\medskip \noindent
FIG. 7. The schematic picture of the excitation process of the 
one orbital wave in the K-edge case. 
$j$ denotes one of the nearest neighbor sites of $i$ 
where $1s \rightarrow 4p$ transition is caused by x-ray. 
The two levels indicate the degenerate $e_g$ orbitals. 
The straight and broken arrows indicate the first and second 
electronic processes in the scattering, respectively. 
\par
\medskip \noindent
FIG. 8. The square of the structure factor for the orbital wave with spin-F 
in the K-edge case: 
$I_K=|F_{\mu \alpha}(\vec q, \vec G)|^2/(2t_0^{-1}|A^{(K)}|^2)^2$. 
(a) $h+k+l$=even, and (b) odd where $(h,k,l)$ is defined in 
a unit cell in the face-centered cubic lattice. The orbital 
state is chosen as $({\pi \over 2} /  {\pi \over 2}+\pi)$. 
The polarization is parallel to the $x$-direction. 
The open and filled circles mean the $(-)$- and $(+)$-modes, respectively. 
\par
\medskip \noindent
Fig. 9 The polarization dependence of the square of the 
structure factor for the orbital wave ((+)-mode) with spin-F 
in the K-edge case. 
(a) $h+k+l$=even, and (b) odd where $(h,k,l)$ is defined in 
a unit cell in the face-centered cubic lattice.
\par
\medskip \noindent
FIG. 10. The square of the structure factor for the orbital wave with spin-A in 
the K-edge case: 
$I_K=|F_{\mu \alpha}(\vec q, \vec G)|^2/(2t_0^{-1}|A^{(K)}|^2)^2$. 
(a) $h+k$=even and $l$=odd, and (b) $h+k$=odd and $l$=odd, 
where $(h,k,l)$ is defined in a unit cell in the tetragonal lattice. 
$J_2/J_1=0.5$ which corresponds to the orbital state 
$(\theta_A/-\theta_A)$ with $\theta_A=1.34$. The polarization is 
parallel to the $x$-direction. 
The open and filled circles mean (a) the $(+-)$- and $(-+)$-modes, 
and (b) the $(--)$- and $(++)$-modes, respectively. 
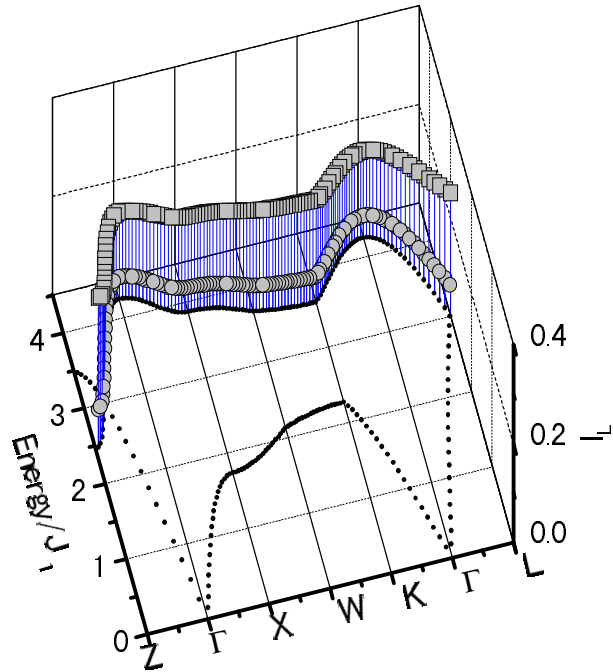
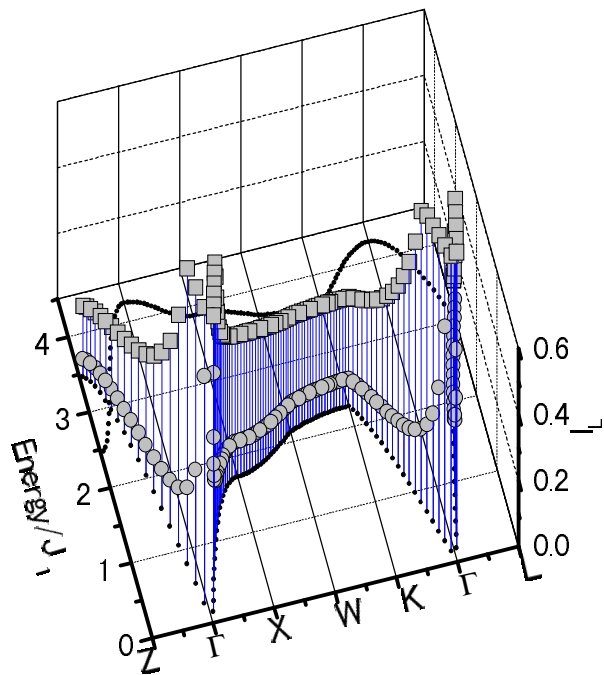
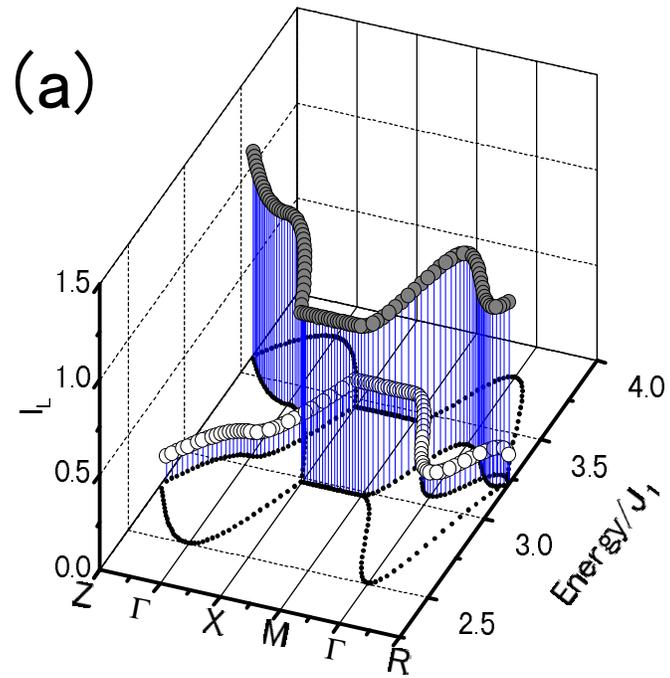
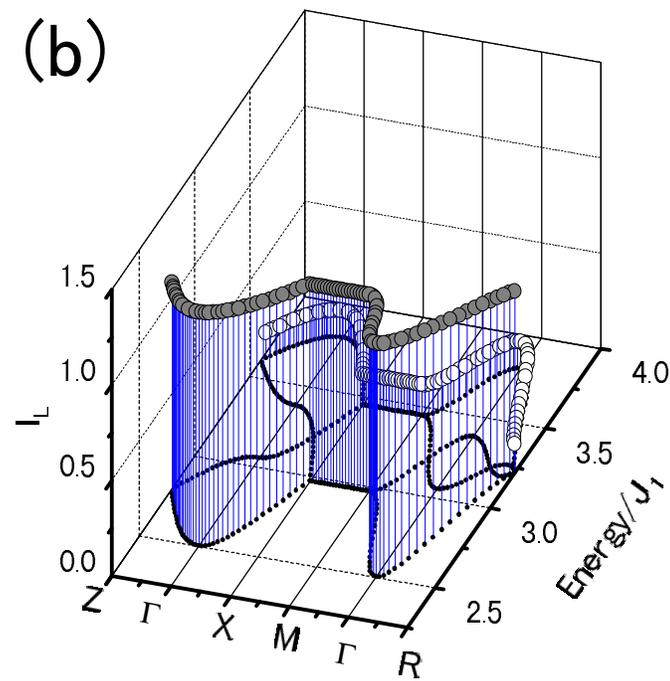
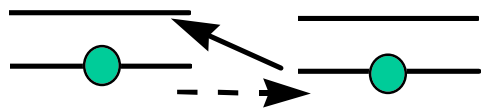
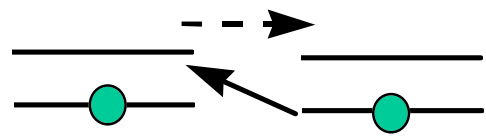
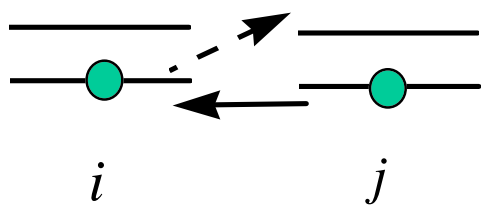
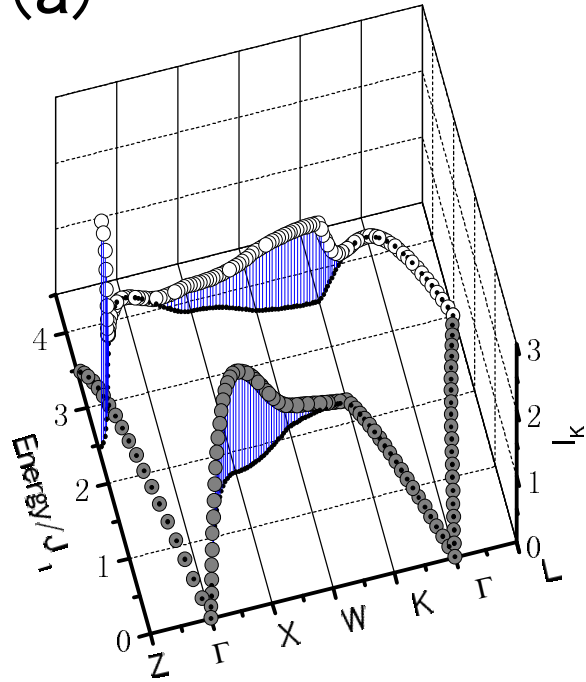
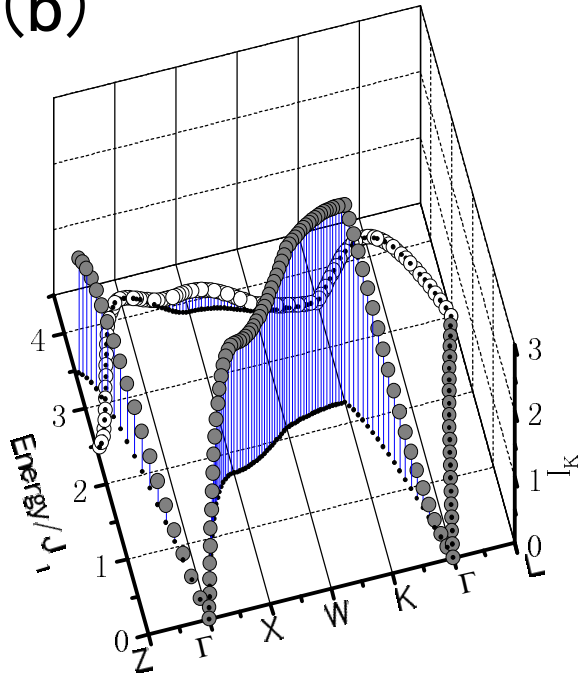
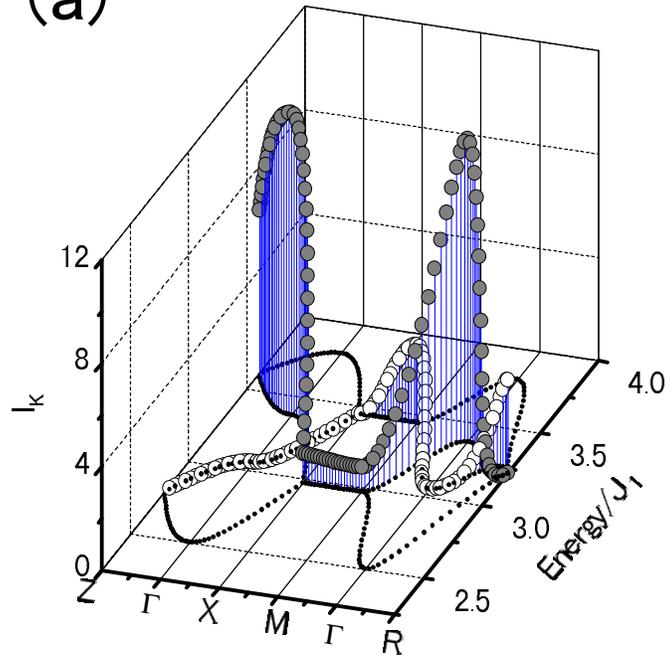
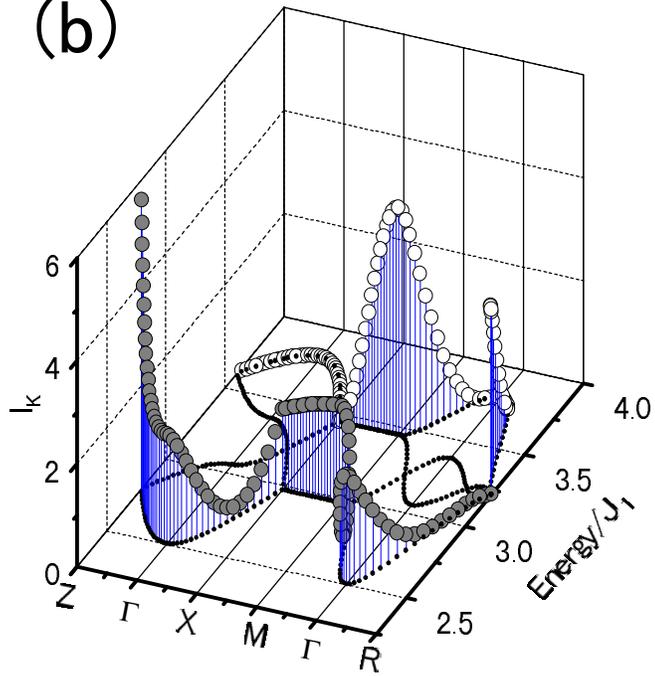
\end{document}